# Reactor monitoring with Neutrinos

Michel Cribier

Astroparticule & Cosmologie
10, rue Alice Domon et Léonie Duquet
F-75205 Paris Cedex 13
France

mcribier@cea.fr

**Abstract**. The fundamental knowledge on neutrinos acquired in the recent years open the possibility of applied neutrino physics. Among it the automatic and non intrusive monitoring of nuclear reactor by its antineutrino signal could be very valuable to IAEA in charge of the control of nuclear power plants. Several efforts worldwide have already started.

## 1. IAEA interest

The International Atomic Energy Agency (IAEA) is the United Nations agency in charge of the development of peaceful use of atomic energy. In particular IAEA is the verification authority of the Treaty on the Nonproliferation of Nuclear Weapons (NPT). To do that jobs inspections of civil nuclear installations and related facilities under safeguards agreements are made in more than 140 states. IAEA uses many different tools for these verifications, like neutron monitor, gamma spectroscopy, but also book keeping of the fuel element composition before and after their use in the nuclear power station. In particular it verify that weapon-origin and other fissile materials that Russia and USA have released from their defence programmes are used for civil application.

Looking for innovative methods, the IAEA ask members states to make a feasibility study to determine whether antineutrino detection methods might provide practical safeguards tools for selected applications. If this method proves to be useful, IAEA has the power to decide that any new nuclear power plants built has to include an antineutrino monitor.

## 2. Physic basis

In a new reactor with normal water the initial fuel consist of enriched uranium rods, with an $^{235}$U content typically at 3.5%, the rest is $^{238}$U. As soon as the reactor is operating, reactions of neutron capture on $^{238}$U produce $^{239}$Pu (and $^{241}$Pu), which then contribute also to the energy production. Nevertheless the net balance in plutonium is positive and a standard pressurized water power reactor produces around 200 kg of plutonium per year.

Every fission of a fissile isotope produce two fissions fragments of unequal masses. The distribution of the lightest fragment is centred around A = 94 for fission of $^{235}$U, and centred around A = 102 in the case of $^{239}$Pu. All these nuclei, too rich in neutrons, are extremely unstable and thus beta decay toward stable nuclei with an average of 6 ß decays and thus with 6 antineutrinos. In these processes several hundreds of unstable nuclei, with their excited states are involved, which makes very difficult to understand details of the physics; moreover, the most energetic antineutrinos, which are detected more easily by the neutrinos detectors, are produced in the very first decays, involving nuclei with typical lifetime much smaller than a second.

|                              | $^{235}$U              | $^{239}$Pu              |
|------------------------------|------------------------|-------------------------|
| released energy per fission  | 201.7 MeV              | 210.0 MeV               |
| Mean energy of ν             | 2.94 MeV               | 2.84 MeV                |
| ν per fission > 1.8 MeV      | 1.92                   | 1.45                    |
| average inter. cross section | $\approx 3.2\ 10^{-43}$ cm$^2$ | $\approx 2.76\ 10^{-43}$ cm$^2$ |

Table. Main characteristics of antineutrinos originating from $^{235}$U and $^{239}$Pu fission

Nevertheless based on predicted and observed ß spectra, the number of antineutrinos per fission from $^{239}$Pu is known to be less than the number from $^{235}$U, and the energy released bigger by 5%. Hence an hypothetical reactor able to use only $^{235}$U would induce in a detector an antineutrino signal 60% higher than the same reactor producing the same amount of energy but burning only $^{239}$Pu (see table 1). This sizeable difference offer a handle to monitor changes in the relative amounts of $^{235}$U and $^{239}$Pu in the core. Merged with the high penetration power of antineutrinos, this provide a new mean to make remote, nonintrusive measurements of plutonium content in reactors [1].

In most of the presently considered detectors, antineutrinos are detected via the inverse beta decay process on quasi-free protons in hydrogenous scintilla tor: $\overline{v}_e + p \rightarrow e^+ + n\ \Box v_e$, with a threshold at 1.8 MeV. The positron and the neutron are detected in a delayed coincidence, allowing strong rejection of the much more frequent singles backgrounds due to natural radioactivity.

Because the antineutrino signal from the reactor decreases as the square of the distance from the reactor to the detector a precise "remote" measurement is really only practical at distances of a few tens of meters if one is constrained to "small" detectors of the order of few cubic meter in size.

## 3. Pioneers in Kurchatov

The potentiality to address certain safeguards applications was recognized long time ago by Mikaelian et al. [2]. The correlation of the antineutrino signal with the thermal power and the burn-up was demonstrated by the Bugey [3] and Rovno experiments [4]. What makes this old idea possible today is our present understanding of the oscillation mechanism which guarantee that the signal recorded by a neutrino detector at less than 200 meters from a reactor is not significantly affected, or could be corrected for. In this respect the results of KamLAND detector [5] as a global monitor of remote ≈ 180 km) power plants is impressive.

## 4. Efforts in the USA

The experimental program for development of nonproliferation detectors in the United States is led by Lawrence Livermore National Laboratory and Sandia National Laboratories. The LLNL/SNL work has consisted of installing and operating a prototype detector at the 3.46 GWth San Onofre Nuclear Generating Station (SONGS) in Southern California. The detector [6], now operating at SONGS at a distance of 24.5 meters from the core in the tendon gallery (fig. 1a), and with an overburden of about 25 m.w.e., is shown in figure 1b. The shielding consists of a muon veto system for rejecting cosmic ray backgrounds, a water/polyethylene shield to reject neutron and gamma backgrounds. The central detector, which registers antineutrino interactions, has a one cubic meter active liquid scintillator doped with gadolinium (0.64 ± 0.06 ton), seen by eight 9" PMTs. The overall footprint including shielding is 2.5 meter × 3 meter.

In this condition the rate predicted at the beginning of the reactor fuel cycle is approximately 3800 ± 440 antineutrino interactions per day for a perfectly efficient detector. The overall efficiency to detect antineutrino interaction via positron neutron delayed coincidence is 10.7% with a signal to background close to 4. The number of antineutrino events observed, 459 ± 16/day is in good agreement with the expected rate deduced from simulation.

Changes in reactor power can quickly (within a few hours) be detected by tracking the antineutrino rate. The plot of daily rate versus time (Figure 2) also shows a two sigma deviation of the antineutrino rate from a constant value over a six month period, with the linear reduction in total rate consistent with a prediction that includes a fuel burn up estimate. Current effort is focused on confirming the indications of fuel burn up seen in this data. Already these results, although modest for neutrino specialists, are convincing enough for external viewers which correlate usually a neutrino detector with a huge apparatus.

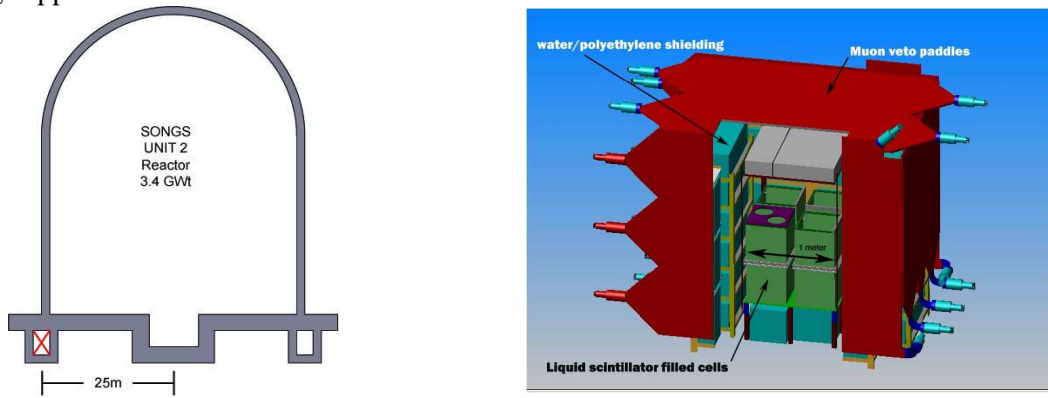

Figure 1. The SONGS detector (right) located in the tendon gallery (left)

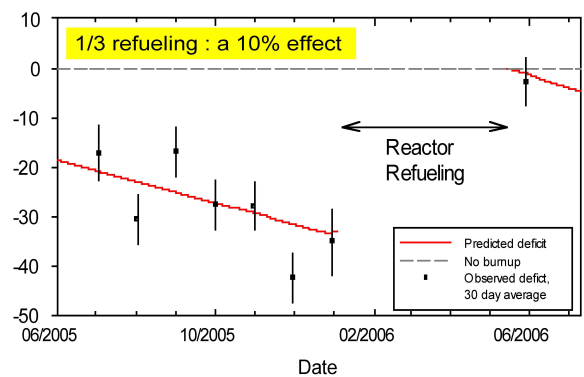

Figure 2. The impact of the refueling is clearly seen on the antineutrino record

## 5. Efforts in France

The Double Chooz collaboration, an experiment [7] mainly devoted to study the fundamental properties of neutrinos, is also in a good position to evaluate the interest of using antineutrino detection to remotely monitor nuclear power station. Indeed, without any extra experimental effort, the near detector of the Double Chooz experiment will provide the most important data set of antineutrino detected ($5 \times 10^5$ ν per year). The precise energy spectrum recorded at a given time will be correlated to the fuel composition and to the thermal power provided by EDF; it is expected that individual component due to fissile element ($^{235}$U, $^{239}$Pu) could be extracted with some modest precision and serve as a benchmark of this techniques.

5.1. Toward a better understanding of the antineutrino spectrum

The IAEA recommends the study of specific safeguards scenarios. Among its concerns are the confirmation of the absence of unrecorded production of fissile material in declared reactors and the monitoring of the burn-up of a reactor core. The time required to manufacture an actual weapon estimated by the IAEA (conversion time), for plutonium in partially irradiated or spent fuel, lies between 1 and 3 months. The significant quantity of plutonium is 8 kg, to be compared with the 3 tons of 235U contained in a Pressurized Water Reactor (PWR) of power 900MWe enriched to 3%. The small magnitude of the expected signal requires a careful feasibility study.

The proliferation scenarios of interest involve different kinds of nuclear power plants such as light water or heavy water reactors (PWR, BWR, Candu...), it has to include isotope production reactors of a few tens of MWth, and future reactors (e.g., PBMRs, Gen IV reactors, accelerator-driven sub-critical assemblies for transmutation, molten salt reactors). To perform these studies, core simulations with dedicated Monte-Carlo codes are being developed in France. The particle transport code MCNPX coupled with an evolution code solving the Bateman equations for the fission products are combined within a package called MURE (MCNP Utility for Reactor Evolution) [8]. It computes accurately the amount of all β- emitters produced during the operation of a nuclear power plant.

To fulfil the goal of nonproliferation, additional laboratory tests and theoretical calculations should also be performed to more precisely estimate the underlying neutrino spectra of plutonium and uranium fission products, especially at high energies. As concluded by P. Huber and Th. Schwetz [9] to achieve this goal a reduction of the present errors on the antineutrino fluxes of about a factor of three is necessary. This is the basis to the important effort to better understand the antineutrino spectrum. More details can be found in S. Cormon contribution [10].

A careful evaluation and propagation of all sources of error is under study in french groups. When building the total spectra, experimental data on the energy and spin and parity of all known nuclear levels are used to determine the shape and error of individual β-branches. In the case of nonunique forbidden β transition, different levels of approximation are used to parameterize the spectrum shape. When computing the amount of β-emitters present in the reactor core at a given time, the effect of the uncertainty on fission yields can be estimated numerically via a limited number of MURE simulations.

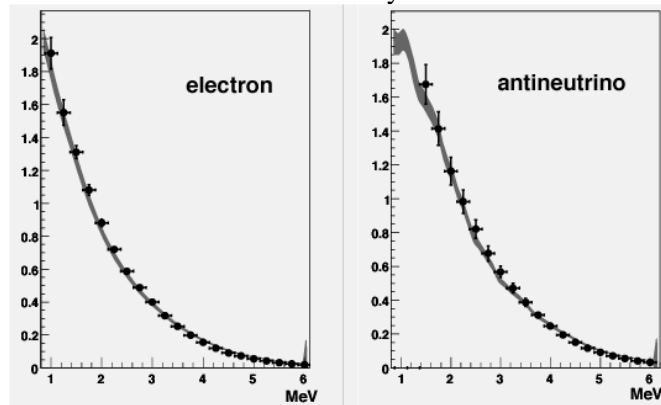

Figure 3. Comparison (Preliminary) with the simulation (shaded area) and Schreckenbach's data [11] for electrons and antineutrino

Thanks to the description of each single β-branch, no extra error is involved when converting the electron total spectrum to an antineutrino spectrum. Once a final error is quoted in each 100 keV bin, this method will allow quantifying for the first time the sensitivity of the neutrino probe to the isotopic composition of the reactor core, that is what minimum change of the $^{235}$U/$^{239}$Pu ratio is detectable. Uncertainties in the normalization and the large amount of common nuclei in the $^{235}$U and $^{239}$Pu decay chains induce large error correlations, which, if treated properly, will increase the sensitivity of the neutrino probe when comparing two spectra.

This promising study will also provide the error budget for an independent measurement of the total nuclear power released by the core. In the low energy part of the spectrum (<4 MeV) one expects the final accuracy to be high enough to provide useful constraints on the total spectrum shape for the neutrino oscillation analysis. At higher energy these simulations can tag few critical poorly known isotopes and trigger an experimental program in order to complete the existing databases.

Finally, the computed energy spectrum of antineutrino can be coupled to a detailed GEANT4 simulation of the Double-Chooz detector or of a small prototype antineutrino detector.

## 6. Toward a prototype of neutrino monitor

If we want to propose to the IAEA a neutrino detector able to help in monitoring future nuclear power plants, the next step in this effort has to merge the two present experimental approaches: the Double Chooz approach with a good energy measurement, a good signal to noise ratio, but expensive and sophisticated; and the SONGS approach with a robust, simple, automatic, cheap, but with poor antineutrino detection efficiency, a modest signal to background ratio, and poor energy resolution. We thus are considering a new prototype with a size small enough to be installed very close to the reactor core (30 meters or so), but using a technique able to clearly sign antineutrinos. Such a prototype will be considered as a demonstrator to be shown to the IAEA and at the same time it is already usable tool to measure the thermal power.

As an intermediate goal, we can foresee measurements with this prototype at ILL with its core of roughly pure $^{235}$U. It would allow the recording of a very pure neutrino signal from $^{235}$U fission only followed by the evolution due to burn-up. Such a clean experiment would help to calibrate the neutrino signal versus the thermal power, and will also give some confidence for the simulation effort.

A brazilian team is currently developing an antineutrino detector to be installed in the close vicinity of a new power plant in Angra [12]. This detector will be used to monitor the reactor activity, and to provide an additional tool on verification of safeguards on Nonproliferation. The planned turnon dates are 2008 for the very near detector and 2013 for the Angra complete configuration.

### 7. Distant monitoring

Neutrinos travels long distance through dense materials without being stopped. Hence it forbids to totally hide some nuclear activities involving fission like a clandestine nuclear power or nuclear test. Moreover, in the case of a test, the unique signature given by the antineutrino interaction, in coincidence with other methods like seismic waves, transform an hint into a proof.

Based on this principal several groups around the world propose network of huge detectors able to detect, locate and measure the power of nuclear reactors and of bombs test. In the mean time a more modest neutrino detector used in coincidence with the standard seismic network could sign unambiguously the real nature of a kton test. This aspect of the use of antineutrinos was covered in the previous proceeding of Neutrino 04 [13] and was the subject of a dedicated workshop [14]. In this approach HanoHano is a good example of such detector [15].

### 8. Neutrinos for Peace

All the applications of our knowledge of neutrinos seems surprising for physicists which, for many years, consider this particle as the most elusive one. It is remarquable that so quickly a very fundamental research could turn into applications: it is even more enjoyable that the first applications envisaged for this unusual particle is the control of arm races and not a new weapon. For all these reasons, I would gladly propose to name these worldwide efforts: Neutrinos for Peace.

**References.**
[1]   A. Bernstein, et. al., J. Appl. Phys. 91 (2002) 4672
[2]   Mikaelian L.A., Proc. Int. Conference Neutrino-77, v. 2, p. 383-387
[3]   Y. Declais et al., Nucl. Phys. B434 (1995) 503
[4]   Yu. V. Klimov, et. al., Atomic Energy, 76 (1994) 123
[5]   K. Eguchi et al., KamLAND Collab., Phys. Rev. Lett. 90 (2003) 021802
[6]   N.S. Bowden et al., ArXiv:physics/0612152
[7]   F. Ardellier et al., ArXiv:hep-ex/0606025
[8]   Mal'plan O. et al. In Proceedings of the ENC 2005 (CD-Rom) (2005) 1-7.
[9]   P. Huber, Th. Schwetz, Phys.Rev. D70 (2004) 053011
[10]  S. Cormon et al., *these Proceedings*
[11]  K. Schreckenbach et al., Phys. Lett. B160 (1985) 325
[12]  J.C. Anjos et al., Nucl.Phys.Proc.Suppl.155:231-232,2006. ; ArXiv: hep-ex/0511059
[13]  J.C. Learned, Nucl. Phys. B (Proc. Suppl.) 143 (2005) 152-156

[14]    Neutrino Science 2005, http://www.phys.hawaii.edu/~sdye/hnsc.html
[15]    S. Dye, ArXiv:hep-ex/0611039